\documentclass[
    preprint,
    nofootinbib,
    amsmath,amssymb,
    aps,
    prd
]{revtex4-2}

\usepackage{graphicx}               
\usepackage{dcolumn}                
\usepackage{bm}                     
\usepackage{ulem}
\usepackage{braket}
\usepackage{mathtools}
\usepackage[colorlinks]{hyperref}   
\usepackage[dvipsnames]{xcolor}

\begin{document}

\preprint{CALT-TH 2025-019}

\title{Correlation functions of von Neumann entropy}

\author{Mathew W.\ Bub}
\email{mbub@caltech.edu}
\author{Allic Sivaramakrishnan}
\email{allic@caltech.edu}

\affiliation{
    Walter Burke Institute for Theoretical Physics \\
    California Institute of Technology, Pasadena, CA 91125, USA
}

\begin{abstract}
In this note, we study two-point correlation functions of modular Hamiltonians. We show that in general quantum systems, these correlators obey properties similar to those of von Neumann entropy and capacity of entanglement, both of which are special cases of these correlators. Then we specialize to two spacelike-separated spherical subregions in conformal field theories. We present direct computations of the vacuum two-point function that confirm its equivalence to the stress-tensor conformal block. We explore the two-point function in various kinematic regimes, including imaginary time separation between subsystems. The material presented in this note may be useful for further studying modular Hamiltonian correlators in generic quantum systems and in conformal field theories, including those with holographic duals.

\end{abstract}

\maketitle

\thispagestyle{empty} 

\newpage
\clearpage 
\setcounter{page}{1}

\tableofcontents

\section{Introduction} \label{sec:intro}

Many entropic measures studied in information theory capture correlations between the spectra of two subsystems. This is useful in quantum systems, where for example von Neumann entropy quantifies entanglement. 

Correlation functions of measurable quantities like local fields also encode correlations between subsystems. In quantum gravity, however, measurements may be more challenging to define due to intrinsic dependence on the nature of the observer and the measurement procedure. For example, see \cite{witten1962gravitation} for general discussion and  \cite{DeVuyst:2024pop, Maldacena:2024spf} for recent work on the observer dependence of entropy. By contrast, von Neumann entropy in conformal field theory (CFT) captures features of quantum gravity in anti-de Sitter space (AdS)  \cite{Maldacena:1997re} in a more universal way. While correlators of local operators often depend heavily on specific properties of those operators, and different theories possess different operators, measures of entropy are defined identically in any finite-dimensional quantum system. Tomita-Takesaki theory provides a rigorous mathematical basis for entropic quantities in infinite-dimensional Hilbert spaces \cite{Takesaki:1970aki}.  Study of von Neumann entropy in the AdS/CFT correspondence \cite{Ryu:2006bv,Hubeny:2007xt,Engelhardt:2014gca} has led to the derivation of a unitary Page curve for evaporating black holes \cite{Penington:2019npb,Almheiri:2019psf}. 

In spite of their differences, CFT correlators and entropic measures can capture similar features of a quantum state. For example, boundary correlators encode realistic bulk observers, which are generically dynamical, massive, and fully interacting, while using technology from Tomita-Takesaki theory, a more universal notion of bulk time can be defined from the boundary \cite{Leutheusser:2021frk,Leutheusser:2022bgi, Doi:2022iyj, Doi:2023zaf}. Relations between the two approaches have also been explored \cite{Jafferis:2020ora, deBoer:2022zps}. The average null energy condition has been derived both from relative entropy \cite{Faulkner:2016mzt} and also causality properties of CFT correlators \cite{Hartman:2016lgu}. As discussed in \cite{Hartman:2016lgu}, there is evidence that the relationship between entanglement and causality may be more general.

In this note, we study objects that may help connect local correlation functions in quantum field theory and entropic measures in quantum information theory. The von Neumann entropy $S = -\text{tr}(\rho \log \rho)$ of density matrix $\rho$ is equivalently written as $S= \braket{K}$, the one-point function of the modular Hamiltonian $K$, which is defined by $\rho = e^{-K}$. However, higher-point correlation functions of $K$ are far less well-studied. We will explore these correlators, for instance $\text{tr}(\rho \log \rho_1 \log \rho_2)=\braket{K_1 K_2}$ with density matrices $\rho_i = e^{-K_i}$. We mainly consider setups in which the $\rho_i$ are reduced density matrices corresponding to distinct subsystems and $\rho$ is the state of the full system. 

We summarize this note here. In Section \eqref{sec:2point}, we give several general properties. We find that $\braket{K_1 \cdots K_n} \geq 0$, that $\braket{K_1 \cdots K_n} = 0$ if and only if at least one $\rho_i$ is a pure state, and that the connected correlator obeys $\braket{K_1 \cdots K_n}_c = 0$ when the entanglement spectrum of at least one $\rho_i$ is flat, which includes pure and maximally mixed states. These properties are analogous to those obeyed by von Neumann entropy and also the capacity of entanglement $C=\braket{K^2}-\braket{K}^2$, which are special cases of the correlators we consider. We conclude that in this sense, correlators of $K_i$ meaningfully generalize von Neumann entropy to higher points. Then in Section \eqref{sec:calculation}, we present direct computations that confirm that $\braket{K_1 K_2}_c$ for ball-shaped regions in vacuum CFT$_d$ is given by the stress-tensor conformal block in a kinematic regime in which equivalence is expected. We also compute this two-point function for subregions separated in imaginary time to provide data for studying analytic continuations to kinematic regimes that are currently unexplored.

One aim of this note is to present properties and explicit details of computations in order to facilitate a more systematic exploration of $\braket{K_1 K_2}_c$. We briefly discuss what studying this quantity further may teach us. 

The relationship between the modular Hamiltonian and the stress-tensor OPE block was first observed in \cite{Czech:2016tqr, Czech:2016xec} and has since been further studied along with OPE blocks in \cite{deBoer:2016pqk,Sarosi:2017rsq,Long:2019fay,Long:2020wey,Long:2020zeq,Nath:2024aqh, Verlinde:2019xfb}. These works find that $\braket{K^2} - \braket{K}^2$ computes $\braket{\mathcal{A}^2}-\braket{\mathcal{A}}^2$, where $\mathcal{A}$ is an operator whose expectation value is the area of the bulk Ryu-Takayanagi surface. This was proven in \cite{DeBoer:2018kvc} in a certain setup and is expected to hold more generally \cite{Faulkner:2017vdd,Jafferis:2015del,Engelhardt:2014gca}.\footnote{Studying correlators of OPE blocks may also furnish geodesic Witten diagrams \cite{Hijano:2015zsa} for timelike separated points.}  As $\braket{K_1 K_2}_c$ correlators for distinct ball-shaped regions are generically finite in the vacuum and reproduce area correlators in the bulk, they may nicely capture how the bulk geometry fluctuates. Higher-order corrections may capture correlations between measures of bulk entanglement \cite{Engelhardt:2014gca}.

Entropic measures are notoriously challenging to compute in CFT in comparison to correlators, and usually the replica trick is used. However, the stress-tensor block is known, so in this simple case extracting the vacuum correlator $\braket{K_1 K_2}_c$ from a four-point function is trivial. Motivated by this, one might investigate whether a more general relationship holds, which may provide a novel way to compute entanglement measures using conformal bootstrap methods. An existing point of comparison is \cite{Balakrishnan:2020lbp}, where the modular Hamiltonian was computed perturbatively to all orders in a source and written entirely in terms of local operators in Lorentzian signature. One may also investigate whether multipoint correlators of local operators capture multipartite entanglement.

Recent work has proposed notions of entanglement-like measures for timelike separations \cite{Doi:2022iyj, Doi:2023zaf, Milekhin:2025ycm}. We expect that the quantity $\braket{K_1 \cdots K_n}_c$ can also be analytically continued from spatial subregions to subregions that are timelike separated. How entanglement measures behave under this type of analytic continuation may be subtle. However, by connecting correlators of $K_i$ to conformal blocks, one may study such continuations via the better-studied analytic properties of correlation functions. When formulated in holographic theories, analytically-continued correlators of $K_i$ may furnish a holographic definition of bulk time.

Area fluctuations have entered recently into the discussion of potential observational signatures of quantum gravity \cite{Verlinde:2019ade, Parikh:2024zmu} and may be relevant to the near-horizon limit of black holes, for example as described in certain cases by Jackiw–Teitelboim gravity \cite{Mertens:2022irh}. As $\braket{K_1 K_2}_c$ is expected to be a gauge-invariant, finite definition of such correlators in AdS/CFT, its further study in holography may be useful. For work on proposed relations between area fluctuations, entanglement capacity, and observable signatures in quantum gravity in interferometers, see \cite{Verlinde:2019xfb, Verlinde:2019ade} and recently \cite{Carney:2024wnp, Aalsma:2025bcg}. For related work, see also \cite{Banks:2021jwj,Banks:2024cqo,Banks:2023wua} and citations therein.

\section{General properties of entropy correlators} \label{sec:2point}

Consider a quantum system $\mathcal{S}$ that can be partitioned into various subsystems $\mathcal{S}_1, \ldots, \mathcal{S}_n$ with finite-dimensional Hilbert spaces $\mathcal{H}$ and $ \mathcal{H}_1, \ldots, \mathcal{H}_n$, respectively. Partial traces of the density matrix $\rho$ yield reduced density matrices $\rho_{1}, \ldots, \rho_{n}$, which describe the states of the respective subsystems. The normalization of $\rho$ is tr$(\rho)=1$. Correlation functions of operators that differ from the identity only in a subsystem can be equivalently evaluated with the full or reduced density matrix. For example, partitioning $\mathcal{S}$ into $\mathcal{S}_1, \mathcal{S}_2$, the expectation value $\braket{\mathcal{O}_1 \otimes \mathbf{I}_{2}} \equiv \text{tr}_{12}(\rho ~\mathcal{O}_1 \otimes \mathbf{I}_{2}) = \text{tr}_1 (\rho_1 \mathcal{O}_1)$, where $\rho_1 = \text{tr}_2 ~ \rho$, operators $\mathcal{O}_i$ act as $\mathcal{O}_i: \mathcal{H}_i \rightarrow \mathcal{H}_i$, and $\mathbf{I}$ is the identity operator.

The negative logarithm of a density matrix defines its modular Hamiltonian, for example $\rho_i = e^{-K_i}$. Both $\rho_i$ and $K_i$ are positive semi-definite operators. The von Neumann entropy $S_i$ of subsystem $\mathcal{S}_i$ is
\begin{equation}
S_i = -\text{tr}(\rho_i \log \rho_i)=\braket{K_i}.
\end{equation}
Von Neumann entropy is non-negative, and is zero if and only if the state is pure, for example $\rho_i = \text{diag}(1,0, \dots,0)$.

The one-point functions of $K_i$ are the well-studied von Neumann entropies, but higher-point functions obey a number of universal properties as well. The expectation value of the subtracted quantity $\Delta K_i \equiv K_i -\braket{K_i}$ is $\braket{\Delta K_i} = 0$, but higher moments of $\Delta K_i$ are non-zero and contain fine-grained information about the state. The second moment is $\braket{(\Delta K_i)^2} = \braket{K^2_i}-\braket{K_i}^2 = \braket{K_i K_i}_c$, where the subscript denotes the connected correlator. This is known as the capacity of entanglement $C_i$ \cite{Yao:2010woi,Schliemann:2011},
\begin{equation}
C_i = \braket{K_i K_i}_c = \text{tr}(\rho_i \log^2 \rho_i) - \left(\text{tr}(\rho_i \log \rho_i)\right)^2.
\end{equation}
Because $C_i$ is a variance, it is also sometimes referred to as varentropy. This quantity is non-negative, $\braket{K_i K_i}_c \geq 0$, because it is the expectation value of the positive semi-definite operator $(\Delta K_i)^2$. Furthermore, $\braket{K_i K_i}_c=0$ if and only if the nonzero eigenvalues $\lambda_{i,a}$ of $\rho_i$ are identical, which is known as flatness of the entanglement spectrum,
\begin{equation}
\braket{K_i K_i}_c = 0~~ \text{iff}~~ \lambda_{i,a} = \lambda ~~\text{or }~~ \lambda_{i,a} = 0 ~~\forall ~~ a, ~~~ \text{where}~~~\rho_i = \text{diag}(\lambda_{i,1}, \ldots, \lambda_{i,d_i}),
\end{equation}
where $d_i = \text{dim} (\mathcal{H}_i)$. Flatness is equivalent to $\rho_i$ being a projector state in some basis: $\rho_i = P^{(n)}/n$ where $(P^{(n)})^2 = P^{(n)}$ is a projector onto an $n \times n$ subspace of $\mathcal{H}_i$, in other words, a matrix with $n$ entries of $1$ and $d_i-n$ entries of $0$ on the diagonal. For example, pure and maximally-mixed density matrices have a flat entanglement spectrum. See \cite{DeBoer:2018kvc} for further review.

The interpretation of $\braket{K_i K_i }_c$ is somewhat subtle. Its thermodynamic analog is specific heat capacity at fixed volume. In \cite{DeBoer:2018kvc}, the AdS dual of capacity of entanglement in holographic CFT was identified in vacuum states to leading order in $G_N$ with the variance of the area of the Ryu-Takayanagi surface. However, in the vacuum state, $\braket{K_i K_i }_c$ is divergent for the standard reason that it is the square of an operator that creates infinite-energy excitations. Therefore while $\braket{K_i K_i}_c$ can meaningfully capture features of extremal surface fluctuations, strictly speaking it is not well-defined. A similar statement applies to von Neumann entropy.

In this note, we study correlation functions of $K_i$ for distinct subsystems, for example
\begin{equation}
\braket{K_1 K_2} = \text{tr}(\rho \log \rho_1 \log \rho_2) = \text{tr}(\rho K_1 K_2),~~~~~~ \braket{K_1 K_2}_c = \braket{K_1 K_2}-\braket{K_1} \braket{K_2}.
\end{equation}
We can think of such quantities as correlation functions of von Neumann entropy for reasons that will become clear. These correlators are a generalization of entanglement capacity, $\braket{K_i K_i}_c$. 

We focus on two-point functions, but similar conclusions apply at higher points. Correlators of $\Delta \mathcal{O} = \mathcal{O}-\braket{\mathcal{O}}$ define connected correlators of $\mathcal{O}$. For example, the connected three-point correlator is $\braket{K_1  K_2  K_3}_c \equiv \braket{\Delta K_1 \Delta K_2 \Delta K_3}$, or
\begin{equation}
\braket{K_1  K_2  K_3}_c 
= \braket{K_1 K_2 K_3} - \braket{K_1}\braket{K_2 K_3}-\braket{K_2}\braket{K_1 K_3}-\braket{K_3}\braket{K_1 K_2}+2\braket{K_1}\braket{K_2}\braket{K_3}.
\end{equation}

To understand what information these correlators of $K_i$ encode, we give some elementary properties of $\braket{K_1 K_2}_c$. In tensor-product states,
\begin{equation}
\rho = \rho_1 \otimes \rho_2: ~~  ~~\braket{K_1 K_2}= \braket{K_1}\braket{K_2} ~~\implies~~ \braket{K_1 K_2}_c = 0.
\end{equation}
In fact, this property is generic for correlators in tensor-product states, as $\braket{\mathcal{O}_1 \mathcal{O}_2}_c = 0$ when $\mathcal{O}_1$ is nontrivial in $\mathcal{H}_1$ but the identity in $\mathcal{H}_2$, and similarly for $\mathcal{O}_2$. Incidentally, for $\rho = \rho_1 \otimes \rho_2$, the mutual information $I_{12} = S_1+S_2-S_{12}$ between subsystems $\mathcal{S}_1, \mathcal{S}_2$ vanishes as well, $I_{12} = 0$. 

Correlators of modular Hamiltonians obey a more non-trivial property, that $\braket{K_1 K_2}_c=0$ when at least one of $\mathcal{S}_1, \mathcal{S}_2$ has a flat entanglement spectrum,
\begin{equation}
\rho_1 = P^{(n_1)}/n_1~~ \text{or} ~~ \rho_2 = P^{(n_2)}/n_2 \implies 
\braket{K_1 K_2}_c = 0.
\label{KKflatnessproperty}
\end{equation}
Note that $\braket{K_1 K_2}_c \neq 0$ implies that the entanglement spectra of $\rho_1$ and $\rho_2$ are not both flat, but also that the state is not a tensor-product state. 

We prove \eqref{KKflatnessproperty} as follows. Suppose that in some basis, $\rho_1$ is a projector state, $\rho_{1,ab} =  P^{(n)}_{ab}/n$. Consider two subsystems $\mathcal{S}_1, \mathcal{S}_2$ with joint density matrix $\rho_{ab;cd}$. Using matrix notation in which repeated indices are not summed over,
\begin{equation}
\rho_{1,ab} = \sum_{cd} \rho_{ab;cd}, ~~~\rho_{2,cd} = \sum_{ab} \rho_{ab;cd},
~~~~
\braket{K_1 K_2} = \sum_{abcd}\rho_{ab;cd}\log \rho_{1,ba} \log \rho_{2,dc}.
\end{equation}
We then have
\begin{align}
\braket{K_1 K_2} &= \sum_{abcd}\rho_{ab;cd} P_{ab}^{(n)} \log (1/n) \log \rho_{2,dc} = \log (1/n) \sum_{\substack {acd \\ a \leq n  }} \rho_{aa;cd}  \log \rho_{2,dc} 
\nonumber
\\
&= \braket{K_1}\braket{K_2}-\log (1/n) \sum_{\substack {acd \\ a > n  }} \rho_{aa;cd}  \log \rho_{2,dc}.
\label{KKFlatSpectrumProof}
\end{align}
We can diagonalize $\rho_{2,dc}$ without loss of generality. Finally, we can show that $\rho_{aa;dd}=0$ for $a > n$ as follows. For $a>n$, $\rho_{1,aa} = P^{(n)}_{aa}/n = 0$, which implies that $\sum_d \rho_{aa;dd} = 0$ for $a>n$. Because $\rho$ is positive semi-definite, its diagonal entries in any basis are non-negative, so the vanishing of the sum implies $\rho_{aa;dd} = 0$ for all $d$ when $a>n$. The second term in \eqref{KKFlatSpectrumProof} therefore vanishes, which completes the proof.

While we focus on $\braket{K_1 K_2}_c$, the full correlator is nonnegative, 
\begin{equation}
\braket{K_1 K_2} \geq 0.
\end{equation}
As evident in \eqref{KKFlatSpectrumProof}, because $\rho, K_1, K_2$ are all positive semidefinite operators and $K_1, K_2$ act on different Hilbert spaces, we can without loss of generality write $\text{tr}(\rho K_1 K_2)$ in a basis that simultaneously diagonalizes $K_1, K_2$, from which it is clear that $\braket{K_1 K_2} \geq 0$.

Another property is that 
\begin{equation}
\braket{K_1 K_2} =0 ~~~ \text{iff} ~~~\rho_1 ~\text{or} ~\rho_2 ~ \text{is pure}. 
\end{equation}
To prove this, first recall that $\braket{K_1 K_2}$ in a basis that simultaneously diagonalizes $\rho_1, \rho_2$ is the sum of positive terms. Therefore, $\braket{K_1 K_2} =0$ if and only if every term in the sum is zero. Every non-zero term in the sum is proportional to a product of the logarithm of eigenvalues,  $\log(\lambda_{1,a})\log(\lambda_{2,b})$, and so every term in the sum is zero if and only if at least one of $\rho_1$, $\rho_2$ is a pure state.

Finally, if we consider $\mathcal{S}_1, \mathcal{S}_2$ to be non-overlapping, spacelike-separated spatial subregions in a quantum field theory, we expect that unlike $\braket{K_1 K_1}_c$, $\braket{K_1 K_2}_c$ lacks any coincident-point or lightcone singularity, and is finite and therefore well-defined. This is indeed the case for ball-shaped regions in the CFT vacuum, where the explicit expressions for the $K_i$ are known. Choosing $\mathcal{S}_i$'s to instead be timelike-separated subregions may lead to some singularity and would be interesting to investigate.

It is clear the properties we have given generalize to higher-point correlators of $K_i$. The properties of these correlators closely resemble those of von Neumann entropy and entanglement capacity, which are special cases of these correlators. We therefore conclude that correlators of $K_i$ can be considered, in this specific sense, to be the higher-point correlation functions of von Neumann entropy.

\section{Computing \texorpdfstring{$\braket{K_1 K_2}_c$}{<K1 K2>c} in CFT$_d$} \label{sec:calculation}

In CFT$_d$, the operator $\Delta K_i \equiv K_i-\braket{K_i}$ for a ball-shaped region in the vacuum state has been identified as the stress-tensor OPE block \cite{Czech:2016tqr, Czech:2016xec} up to an overall numerical coefficient. In this section, we explore this relation in a variety of ways. We provide direct computations of $\braket{K_1 K_2}_c$ in the vacuum of CFT$_d$ and compare to the stress-tensor conformal block. We also explore previously uncharted kinematic regimes, namely when the subregions are separated in imaginary time.

We now state our setup. Subsystems $\mathcal{S}_1$, $\mathcal{S}_2$ are each codimension-1 balls of radius $L$. In the vacuum of a Lorentzian CFT, the operator $\Delta K$ admits a local expression in terms of the stress tensor \cite{Casini:2011kv},
\begin{equation} \label{eq:K}
    \Delta K = 2\pi \int_{B_L} d^{d-1}\bm{x} \, \frac{L^2 - \bm{x}^2}{2L} T_{00}(x),
\end{equation}
where $B_L$ denotes a $(d-1)$-ball of radius $L$ centered at the origin, and boldface denotes a spatial vector, i.e., $x = (x^0, \bm{x})$. This expression is obtained via a conformal transformation from the Rindler wedge, for which the vacuum modular Hamiltonian is simply the boost generator \cite{Bisognano:1975,Bisognano:1976}. Beyond these and other simple choices of subregions, the explicit form of the modular Hamiltonian is unknown.

Suppose that we have two spheres whose centers are separated by a spacetime vector $x$. Denote by $K(0)$ the modular Hamiltonian for a ball centered at the origin, and by $K(x)$ the same for a ball centered at $x = (x^0, \bm{x})$. We wish to compute the connected correlator $\braket{K(x) K(0)}_c$. Using \eqref{eq:K},
\begin{equation} \label{eq:KK}
    \braket{K(x)K(0)}_c =  4\pi^2 \int_{B_{\mathrlap{L}}} d^{d-1} \bm{y} \int_{B_{\mathrlap{L}}} d^{d-1} \bm{z} \left( \frac{L^2 - \bm{y}^2}{2L} \right) \left( \frac{L^2 - \bm{z}^2}{2L} \right) \braket{T_{00}(x^0, \bm{x} + \bm{y}) T_{00}(0, \bm{z})}.
\end{equation}
The stress-tensor two-point function is
\begin{equation} \label{eq:TT}
    \braket{T_{00}(x) T_{00}(0)} = \frac{C_T}{d} \left( \frac{(d-1)x^4 - 4d x^2 x_0^2 + 4dx_0^4}{x^{2(d+2)}} \right),
\end{equation}
where $C_T$ is a dimensionless constant \cite{Osborn:1993cr}. We will denote the spatial separation between the centers of the spheres by $a \equiv |\bm{x}|$. While \eqref{eq:K} is derived in Lorentzian signature, we will explore this correlator in the complex $x^0$ plane. When in Euclidean signature, we denote the Euclidean time separation by $\tau$ rather than $x^0$. In other words, to pass between Euclidean and Lorentzian times, we analytically continue the time coordinates of the center of the spheres. We will consider $d = 2$ and $d > 2$ separately.

\subsection{\texorpdfstring{$d=2$}{d = 2}} \label{sec:2d}

\noindent
We study $\braket{K(x)K(0)}_c$ in $d=2$, working first in Euclidean signature. Here, the expression in \eqref{eq:KK} reduces to
\begin{equation}
    \braket{K(x)K(0)}_c = 4\pi^2 \int_{-L}^L dy \int_{-L}^L dz \left( \frac{L^2 - y^2}{2L} \right) \left( \frac{L^2 - z^2}{2L} \right) \braket{T_{00}(\tau, a + y) T_{00}(0, z)},
\end{equation}
which can be evaluated directly. As an example, let us first consider the case of equal times, $\tau = 0$, and non-overlapping subregions. This is trivially equivalent to fixing the two spheres at spacelike separation in Lorentzian signature, $x^0 = 0$ and $a > 2L$. In the equal-time case,
\begin{equation} \label{eq:KK_equal_time_2d}
\begin{aligned} 
    \braket{K(0,a)K(0)}_c &= \frac{\pi^2 C_T}{2L^2} \int_{-L}^L dy \int_{-L}^L dz \, \frac{(L^2 - y^2)(L^2 - z^2)}{(y-z+a)^4} \\
    &= \frac{\pi ^2 C_T}{6} \left[\left(2 - \frac{a^2}{L^2} \right) \log \left(1 - \frac{4L^2}{a^2} \right)-4\right].
\end{aligned}
\end{equation}
It is also possible to directly evaluate the two-point function in a more general case, with $\tau \neq 0$. The result can be expressed compactly in terms of the complex quantities $z \equiv 4L^2/(a - i\tau)^2$ and $\bar{z} = z^*$ as
\begin{equation} \label{eq:KK_2d}
    \braket{K(x)K(0)}_c = \frac{\pi^2 C_T}{6} \left[ \left( 1 - \frac{2}{z} \right) \log (1-z) + \left( 1 - \frac{2}{\bar{z}} \right) \log (1-\bar{z})-4 \right],
\end{equation}
which manifestly reduces to \eqref{eq:KK_equal_time_2d} upon setting $\tau = 0$. We will see shortly that the quantities $z, \bar{z}$ are precisely the appropriate values of the conformal cross ratios that appear in the conformal block.

Let us now compare this result to the stress tensor conformal block. In 2D, the $s$-channel conformal block $g_{\Delta,J}(z,\bar{z})$ with scaling dimension $\Delta$, spin $J$, and identical external operators is given by \cite{Dolan:2000ut, Dolan:2003hv}
\begin{equation} \label{eq:conformal_block_2d}
    g_{\Delta,J}(z,\bar{z}) = \frac{ k_{\Delta + J}(z) k_{\Delta - J}(\bar{z})+k_{\Delta - J}(z) k_{\Delta + J}(\bar{z})}{1 + \delta_{J,0}},
\end{equation}
where
\begin{equation}
    k_\beta(z) = z^{\beta/2} {}_2F_1(\beta/2, \beta/2, \beta, z),
\end{equation}
and $z$ is the usual cross ratio $z = z_{12} z_{34} / z_{13} z_{24}$, with complex coordinates $z_i = x_i - i\tau_i$ encoding the points $(\tau_i, x_i)$. See \cite{Qiao:2020bcs} for a detailed discussion of various coordinate choices and their regimes of validity. Note that conformal blocks with pairwise-identical external scalar operators are independent of the external operator dimensions.

The quantity $\Delta K$ is proportional to the stress tensor OPE block appearing in the OPE of $\mathcal{O}(z_1,\bar{z}_1) \mathcal{O}(z_2,\bar{z}_2)$ in a specific Lorentzian kinematic configuration \cite{Czech:2016tqr, Czech:2016xec}. The OPE block $\mathcal{B}^{\mathcal{O} \mathcal{O}}_{\mathcal{O}_p}$ for primary operator $\mathcal{O}_p$ is defined by repackaging the sum over global descendants in the OPE as
\cite{Czech:2016xec}
\begin{equation}
\mathcal{O}(y_1) \mathcal{O}(y_2) = \frac{1}{(y_1-y_2)^{2\Delta}}\sum_{\mathcal{O}_p} C_{\mathcal{O}\mathcal{O}\mathcal{O}_p}\mathcal{B}^{\mathcal{O} \mathcal{O}}_{\mathcal{O}_p}(y_1,y_2), 
\end{equation} 
with OPE coefficient $C_{\mathcal{O}\mathcal{O}\mathcal{O}_p}$. In Lorentzian signature, the domain of dependence of the $S^{d-1}$ is a causal diamond, and the equivalence between the OPE block and $\Delta K$ occurs when the operator locations $y_1$ and $y_2$ are at the top and bottom tips of this diamond. We therefore need to evaluate the stress-tensor conformal block with $z_i, \bar{z}_i$ analytically continued so that the corresponding points  are the top and bottom tips of the relevant diamonds.

We will begin with the Euclidean configuration
\begin{equation} \label{eq:zi}
    z_1 = -L, \quad z_2 = L, \quad z_3 = a-L- i\tau, \quad z_4 = a +L - i\tau .
\end{equation}
Setting $\tau=0$ is equivalent to inserting the four operators at the same Lorentzian time but different spatial locations. We will however keep the Euclidean time nonzero, $\tau \neq 0$, in order to explore a more general equivalence.

The operator locations \eqref{eq:zi} yield the cross ratio $z = 4L^2/(a-i\tau)^2$. The stress tensor conformal block $g_{2,2}(z, \bar{z})$ is given by 
\begin{equation} \label{eq:2dblock}
    g_{2,2}(z,\bar{z}) = 6\left[ \left( 1 - \frac{2}{z} \right) \log (1-z) + \left( 1 - \frac{2}{\bar{z}} \right) \log (1-\bar{z})-4 \right].
\end{equation}
Next, to make contact with the relation between the modular Hamiltonian and OPE block in \cite{Czech:2016tqr, Czech:2016xec}, we need to continue $z_i$ from this spacelike configuration to one where $\mathcal{O}(z_1,\bar{z}_1), \mathcal{O}(z_2,\bar{z}_2)$ and $\mathcal{O}(z_3,\bar{z}_3), \mathcal{O}(z_4,\bar{z}_4)$ are pairwise timelike separated. This amounts to analytically continuing the block around $z, \bar{z}=0$. However, expanding the block around $z, \bar{z} = 0 $ shows it does not cross any branch cut during this continuation. The block is therefore unchanged under this continuation, and we can simply Wick rotate to the timelike configuration with impunity. The final cross ratios $z, \bar{z}$ are, coincidentally,  the same before and after continuation. Therefore the equality of \eqref{eq:KK_2d} and \eqref{eq:2dblock} proves the equivalence between $\Delta K$ and the OPE block at the level of two-point functions. Both the block and the shadow block satisfy the same conformal Casimir equation, and this was initially used to prove the equivalence of $\Delta K$ and the OPE block \cite{Czech:2016tqr,Czech:2016xec}, but our direct computation shows that in this setup, the stress-tensor conformal block is indeed equivalent to $\braket{K(x)K(0)}_c$ with a constant of proportionality of $
\pi^2 C_T / 36$. For $C_T = 1$, this agrees with the constant of proportionality between the OPE block and the modular Hamiltonian found in \cite{Czech:2016tqr}.

\subsection{\texorpdfstring{$d>2$}{d > 2}} \label{sec:general_d}
In higher dimensions, we will evaluate the expression \eqref{eq:KK} in two tractable regimes: for large spatial separations $a \gg L, \tau$ and for spatially-overlapping balls at different Euclidean times, $a = 0$, $\tau \neq 0$. Note that this is an analytic continuation in time of the more familiar Lorentzian configuration, which is recovered by setting $\tau=0$. 

For $a \gg L$, the integral can be easily evaluated as a series in $a$. For example, in $d=4$ we have that
\begin{equation} \label{eq:KK_4d_series}
    \braket{K(x)K(0)}_c = \frac{16 \pi^4 C_T L^8}{75 a^8} \left[ 1 +\frac{4 \left(6L^2-7 \tau ^2\right)}{3 a^2}+ \frac{2 \left(160 L^4-420 L^2 \tau ^2+147 \tau ^4\right)}{7 a^4} + \mathcal{O}(a^{-6}) \right].
\end{equation}

In the case of the spatially-overlapping subregions, $a = 0$, $\tau \neq 0$, we can obtain a closed form expression by evaluating the integral. We begin by applying the following change of variables to \eqref{eq:KK},
\begin{equation} \label{eq:cov}
    \bm{u} = \bm{y} + \bm{z}, \qquad \bm{v} = \bm{y} - \bm{z}.
\end{equation}
To determine how the region of integration transforms under this change of variables, first notice that $\bm{v}$ may take any value in a ball of radius $2L$. We wish to determine the permitted values of $\bm{u}$ given a fixed $\bm{v} \in B_{2L}$. From \eqref{eq:cov} and using that $\bm{y},\bm{z} \in B_L$, we have the conditions
\begin{equation}
    \bm{y}^2 = \left( \frac{\bm{u} + \bm{v}}{2} \right)^2 \leq L^2, \quad \bm{z}^2 = \left( \frac{\bm{u} - \bm{v}}{2} \right)^2 \leq L^2.
\end{equation}
These conditions require that $\bm{u} \pm \bm{v}$ take values in a ball of radius $2L$. Equivalently, the conditions require that $\bm{u}$ takes values in the region $\mathcal{U}_{\bm{v}} \equiv B_{2L}(\bm{v}) \cap B_{2L}(\bm{-v})$, the intersection of two balls of radius $2L$ centered at $\pm \bm{v}$. We then have
\begin{multline} \label{eq:KK_transformed}
    \braket{K(\tau,0) K(0)}_c = \frac{\pi^2 C_T}{2^{d-1}dL^2} \int\limits_{\mathclap{B_{2L}}} d^{d-1} \bm{v} \int\limits_{\mathclap{\mathcal{U}_{\bm{v}}}} d^{d-1} \bm{u} \left( L^2 - \left(\frac{\bm{u} + \bm{v}}{2}\right)^2 \right) \left( L^2 - \left(\frac{\bm{u} - \bm{v}}{2}\right)^2 \right) \\
    \times \left[ \frac{(d-1)(\bm{v}^2 + \tau^2)^2 - 4d \tau^2 (\bm{v}^2 + \tau^2) + 4d \tau^4}{(\bm{v}^2 + \tau^2)^{d+2}} \right].
\end{multline}

Since the region $\mathcal{U}_{\bm{v}}$ is axisymmetric about $\bm{v}$, let us adopt (hyper-)cylindrical coordinates $(\rho, z, \Omega_{\bm{u}})$ for $\bm{u}$ with the axis aligned along the direction of $\bm{v}$. We write
\begin{equation}
    \bm{u} = \rho \hat{\bm{v}}_\perp + z \hat{\bm{v}},
\end{equation}
where $\hat{\bm{v}}$ is the unit vector in the direction of $\bm{v}$, and $\hat{\bm{v}}_\perp$ is a unit vector orthogonal to $\bm{v}$. The vector $\hat{\bm{v}}_\perp$ is implicitly a function of the angular coordinates $\Omega_{\bm{u}}$ on a $(d-3)$-sphere. We also adopt spherical coordinates $(r, \Omega_{\bm{v}})$ for $\hat{\bm{v}}$, writing
\begin{equation}
    \bm{v} = r \hat{\bm{v}}.
\end{equation}
Here, $\Omega_{\bm{v}}$ are the coordinates on a $(d-2)$-sphere. In these coordinates, we have the relations
\begin{equation}
    \bm{u}^2 = \rho^2 + z^2, \quad \bm{v}^2 = r^2, \quad \bm{u} \cdot \bm{v} = rz.
\end{equation}
Since the integrand in \eqref{eq:KK_transformed} depends on $\bm{u}$ and $\bm{v}$ only through $\bm{u}^2$, $\bm{v}^2$, and $\bm{u} \cdot \bm{v}$, the angular integrals merely give two area factors. We can therefore write \eqref{eq:KK_transformed} in the simple form
\begin{equation} \label{eq:KK_simplified}
\begin{aligned}
    \langle \Delta K(\tau) \Delta K(0) \rangle = \frac{\Omega_{d-2} \Omega_{d-3} \pi^2 C_T}{2^{d-2}dL^2} &\int\limits_0^{2L} r^{d-2} dr \int\limits_0^{\mathclap{2L-r}} dz \int\limits_0^{\mathrlap{\sqrt{4L^2 - (r + z)^2}}} \rho^{d-3} d\rho \\
    &\times \left( L^2 - \frac{1}{4}\left(\rho^2 + (r+z)^2\right) \right) \left( L^2 - \frac{1}{4}\left(\rho^2 + (r-z)^2\right) \right) \\
    &\times \left[ \frac{(d-1)(r^2 + \tau^2)^2 - 4d \tau^2 (r^2 + \tau^2) + 4d \tau^4}{(r^2 + \tau^2)^{d+2}} \right],
\end{aligned}
\end{equation}
where $\Omega_{d-2}$ and $\Omega_{d-3}$ are respectively the areas of the unit $(d-2)$- and $(d-3)$-sphere,
\begin{equation}
    \Omega_n = \frac{2\pi^\frac{n+1}{2}}{\Gamma(\frac{n+1}{2})},
\end{equation} 
and we also used the fact that the integrand is even in $z$ to simplify the limits of integration. The expression \eqref{eq:KK_simplified} can then be easily evaluated in any given dimension. For example, in $d=4$,
\begin{equation}
    \braket{K(\tau, 0) K(0)}_c = \frac{\pi^4 C_T}{60} \left[ \frac{8L^4}{4L^2 \tau^2 + \tau^4} - 3 \left( 1 + \frac{\tau^2}{2L^2} \right) \log \left( 1 + \frac{4L^2}{\tau^2} \right) + 6\right].
\end{equation}

Now, we again compare our computations to the stress tensor conformal block. In $d=4$, for instance, the conformal blocks are given by \cite{Dolan:2000ut,Dolan:2003hv}
\begin{equation}
    g_{\Delta,J}(z, \bar{z}) = \frac{z \bar{z}}{\bar{z} - z} [k_{\Delta - J - 2}(z) \, k_{\Delta + J}(\bar{z}) - k_{\Delta + J}(z) \, k_{\Delta - J - 2}(\bar{z})],
\end{equation}
which gives the stress tensor conformal block
\begin{equation}
    g_{4,2}(z, \bar{z}) = 30 \left[ \frac{z^2(\bar{z}^2-6 \bar{z} + 6 )}{z \bar{z}(z - \bar{z})} \log(1-z) + \frac{\bar{z}^2(z^2 -6 z + 6)}{z \bar{z}(\bar{z} - z)} \log(1-\bar{z}) + 6 \right].
\end{equation}
Now we compare this result with the expression \eqref{eq:KK_4d_series} in the $a \gg L,\tau$ limit. Substituting \eqref{eq:zi} for the cross ratio and expanding, the stress tensor block becomes
\begin{equation} \label{eq:block_4d_series}
    g_{4,2}(z, \bar{z}) = \frac{768 L^8}{a^8} \left[ 1 +\frac{4 \left(6L^2-7 \tau ^2\right)}{3 a^2}+ \frac{2 \left(160 L^4-420 L^2 \tau ^2+147 \tau ^4\right)}{7 a^4} + \mathcal{O}(a^{-6}) \right],
\end{equation}
which is precisely \eqref{eq:KK_4d_series} when setting $\tau =0$ in both cases. Notably, the two are also equal for $\tau \neq 0$, which is outside the standard Lorentzian regime. We leave exploring equality in the full complex time plane to future work. The edge of the wedge theorem may be useful for this purpose. 

\section{Acknowledgments}
We thank Cesar Agon, Alexey Milekhin, and Sridip Pal for comments on draft. We also thank Kathryn Zurek for discussions and suggesting this topic to us. M.B. and A.S. are supported by the Heising-Simons Foundation “Observational Signatures of Quantum Gravity” collaboration grant 2021-2817, the U.S. Department of Energy, Office of Science, Office of High Energy Physics, under Award No. DE-SC0011632, and the Walter Burke Institute for Theoretical Physics. M.B. is also supported by GQuEST funding provided from the U.S. Department of Energy via FNAL: DE-AC02-07CH11359. M.B. acknowledges the support of the Natural Sciences and Engineering Research Council of Canada (NSERC), [funding reference number PGS D- 578032- 2023].

\bibliography{refs}

\end{document}